\documentclass[letter]{aa}     

\usepackage{graphicx}
\usepackage{txfonts}
\usepackage{lipsum}
\usepackage{subcaption}         
\usepackage{lscape}             
\usepackage{placeins}           
                                
\usepackage{hyperref}

\begin{document}
\nolinenumbers
\makeatletter
\renewcommand\makeLineNumber{} 
\makeatother
   \title{The Quiescent Merging Nature of the Coma Cluster Revealed by ICM Velocity Structure}


%

   \author{E. Gatuzz\inst{1},   
           J. Sanders\inst{1}, 
           A. Liu\inst{1,2}, 
           A. Fabian\inst{3}, 
           C. Pinto\inst{4},
           D. Eckert\inst{5}  \and
           S. Walker\inst{6} 
          } 
 
   \institute{Max-Planck-Institut f\"ur extraterrestrische Physik, Gie{\ss}enbachstra{\ss}e 1, 85748 Garching, Germany\\
              \email{egatuzz@mpe.mpg.de}
         \and
         Institute for Frontiers in Astronomy and Astrophysics, Beijing Normal University, Beijing 102206, China
         \and
             Institute of Astronomy, Madingley Road, Cambridge CB3 0HA, UK
         \and
              INAF - IASF Palermo, Via U. La Malfa 153, I-90146 Palermo, Italy 
          \and
             Department of Astronomy, University of Geneva, Ch. d\rq Ecogia 16, CH-1290 Versoix, Switzerland                
          \and
             Department of Physics and Astronomy, University of Alabama in Huntsville, Huntsville, AL 35899, USA                   
             }

   \date{Received November 11, 2025}  

  \abstract  
{ The hot gas permeating galaxy clusters-the intracluster medium (ICM)-is a key tracer of their assembly history and internal dynamics. 
Understanding the motion of this gas provides critical insight into processes such as mergers, turbulence, and energy dissipation in the largest gravitationally bound structures in the Universe. 
The Coma cluster is a nearby, massive system long suspected to be dynamically disturbed. 
Previous high-resolution X-ray spectroscopy with the XRISM mission revealed bulk motions in the cluster core and southern regions. 
Here we present new XRISM Resolve observations of a northern region in Coma, which reveal a coherent velocity gradient of nearly  $530\rm\: $~km/s across the cluster from south to north. 
We find that the hot gas in this northern region exhibits modest line-of-sight motions and uniform thermodynamic properties, indicating relatively mild local disturbances. 
The consistent levels of turbulence throughout the cluster suggest that the energy from a past merger has been distributed on large scales. 
These findings provide compelling evidence for an off-axis merger event and demonstrate how high-resolution X-ray spectroscopy can uncover subtle dynamical signatures in the ICM, offering important constraints for simulations of cluster evolution. 
}

\titlerunning{The Coma cluster ICM velocity structure}
\authorrunning{Gatuzz et al.}
   \maketitle

\section{Introduction}  
Galaxy clusters grow through mergers and accretion, processes that drive turbulent motions and bulk flows in the intracluster medium (ICM), influencing pressure support, metal transport, and energy redistribution \citep{hei10,sch17,zuh18,vaz21}. 
Over the past decade, indirect constraints from X-ray and Sunyaev-Zel’dovich surface-brightness fluctuations have provided statistical insights into the amplitude and spectrum of ICM motions in nearby clusters, including Coma \citep{chu12,zhu18}. 
Direct velocity measurements, however, have been limited by spectral resolution and calibration uncertainties \citep{zhu14,pin15,ang18}, until {\it Hitomi} resolved line-of-sight motions in Perseus, revealing remarkably low turbulence despite strong AGN activity \citep{hit16}. 

The launch of XRISM in 2023 now enables high-resolution spectroscopy across cluster-wide scales. 
Observations of Abell 2029 show a nearly quiescent ICM with low velocity dispersion ($169 \pm 10$~km/s) and a non-thermal pressure fraction of $\sim2.6\%$, while Centaurus exhibits modest bulk flows ($130-310$~km/s) with similarly low turbulence \cite{xri25a,xri25b}. 
The Coma cluster ($z = 0.02333$, \cite{bil18}), a massive nearby cluster undergoing complex mergers, presents a unique environment to study large-scale gas motions. 
It exhibits multiple signatures of ongoing dynamical activity: a disturbed X-ray morphology, a giant radio halo, and substructures associated with the dominant galaxies NGC 4874 and NGC 4889, which exhibit a velocity offset of $\sim 700\rm\: $~km/s \cite{fit87,bri92,whi93,ada05}. 
Infalling groups such as NGC 4839 and NGC 4911/4921 further complicate the cluster's velocity field \cite{ada05}. 
Recen XRISM observations of the core and southern regions revealed substantial bulk motions alongside low velocity dispersions, indicating subsonic turbulence and a dynamically evolved state \cite{xri25c}. 
Here we present new XRISM Resolve observations of a northern region, extending coverage beyond the Performance Verification pointings. 

\vspace{-8mm} 
\section{XRISM  Data reduction}\label{sec_dat}
\vspace{-2mm} 
The northern Coma cluster was observed with {\it XRISM}/Resolve on 2025 January 14 (ObsID 201114010) for a net exposure of 144.8~ks after standard filtering. 
The energy resolution was $4.51 \pm 0.02$~eV at 5.9~keV, corresponding to a velocity uncertainty $<10$~km/s, well below the statistical errors. 

Spectra were extracted for the full field of view and for four quadrants (NE, NW, SE, SW) to probe spatial variations. 
Non-X-ray background contributed $<8\%$ of the counts, and ARFs were computed using {\it Chandra} 2--7~keV images with point sources removed. 
PSF scattering effects were modeled with {\tt xrtraytrace}. 
For further details on the data reduction, including event selection, response generation, and quadrant extraction, see Appendix~\ref{app:data}.

\begin{table*} 
\centering
\footnotesize
\caption{Best-fit parameters in the 2--10 keV energy band for all regions analyzed.}\label{tab_best_fit_models}%
\begin{tabular}{@{}lccccccc@{}}
\hline
\hline
Region  & Temperature & Abundance &  \multicolumn{2}{c}{Redshift} & Turbulent Velocity &  $norm$  & C-stat/dof \\
 & (keV) & (Solar)& $z$ & $v_{\rm bulk}$ ($\rm $~km/s) & $\sigma_{v}$ ($\rm $~km/s)&    & \\
\hline
FOV&$8.14 \pm 0.36$& $0.28 \pm 0.03$ & $0.02273\pm 0.00009$ &$-199\pm 26$ & $167\pm 39$  & $0.195\pm 0.004$& $ 1779/1874 $ \\  
NE& $8.55 \pm 0.83$& $0.38 \pm 0.08$ & $0.02288\pm 0.00030$ &$-155\pm 87$ & $444\pm 256$ & $0.183\pm 0.009$& $ 1017/1261 $\\
NW& $8.20 \pm 0.73$& $0.25 \pm 0.06$ & $0.02258\pm 0.00019$ &$-243\pm 55$ & $118\pm 88 $ & $0.185\pm 0.009$& $ 1114/1287 $\\
SE& $7.60 \pm 0.62$& $0.28 \pm 0.05$ & $0.02257\pm 0.00028$ &$-246\pm 82$ & $303\pm 141$ & $0.101\pm 0.009$& $ 1154/1360 $\\
SW& $8.05 \pm 0.65$& $0.27 \pm 0.06$ & $0.02265\pm 0.00020$ &$-223\pm 58$ & $150\pm 85 $ & $0.183\pm 0.008$& $ 1091/1308 $\\ 
\hline
\hline
\multicolumn{6}{l}{$v_{\rm bulk}$ calculation in $\rm $~km/s includes heliocentric correction. }
\end{tabular}  
\vspace{-3mm}
\end{table*}  

\section{Spectral Fitting} \label{sec_fits} 
\vspace{-1mm}
Spectral fitting was performed in the 2--10~keV band using {\it xspec}~v12.14.1 and Cash statistics \citep{cas79}. 
The ICM was modeled with a single-temperature {\tt bapec} component including thermal and turbulent broadening. 
Redshift, temperature, abundance (proto-solar, \citealt{lod09}), velocity dispersion, and normalization were free parameters. 
Galactic absorption was included via {\tt tbabs} with $N_{\rm H}=8.66\times10^{20}$~cm$^{-2}$ \citep{wil13}, and all velocities were corrected to the heliocentric frame. 
To test for multi-temperature structure, two-temperature ({\tt bapec+bapec}) and log-normal ({\tt blognorm}, \citealt{gat22b}) models were also applied. 
The non-X-ray background (NXB) contributed $<8\%$ of counts and had negligible effect on velocity measurements. 
For full technical details, including model configurations, parameter tying, and MCMC sampling, see Appendix~\ref{app:spectral}.

\vspace{-4.5mm} 
\section{Results and discussion}\label{sec_dis} 

\subsection{Complete Field of View analysis} 
We performed spectral fitting over the full {\it XRISM} Resolve field of view using three ICM models: a single-temperature {\tt bapec}, a two-temperature {\tt bapec+bapec}, and a log-normal temperature distribution with {\tt blognorm}.
To test whether complex models were statistically warranted, we used a simulation-based $\Delta cstat$ method following \citet{buc23}.
For each case, 1000 spectra were simulated from the simpler model, and the 99th percentile of the resulting $\Delta cstat$ distribution defined the critical threshold ($\Delta cstat_{\rm crit}$) corresponding to a 1$\%$ false-positive rate.
We found $\Delta cstat_{\rm crit}=6.38$ for {\tt 2-bapec} and 3.75 for {\tt blognorm}.
Neither model exceeded these thresholds, confirming that a single-temperature component adequately describes the data.
The {\tt blognorm} model only yielded an upper limit on the width parameter, and the second component in the two-temperature fit was consistent with zero.
This supports a uniform thermal structure, consistent with previous findings in the Coma outskirts \citep{arn01,san20,xri25c}.

\begin{figure}        
\centering 
\includegraphics[width=0.45\textwidth]{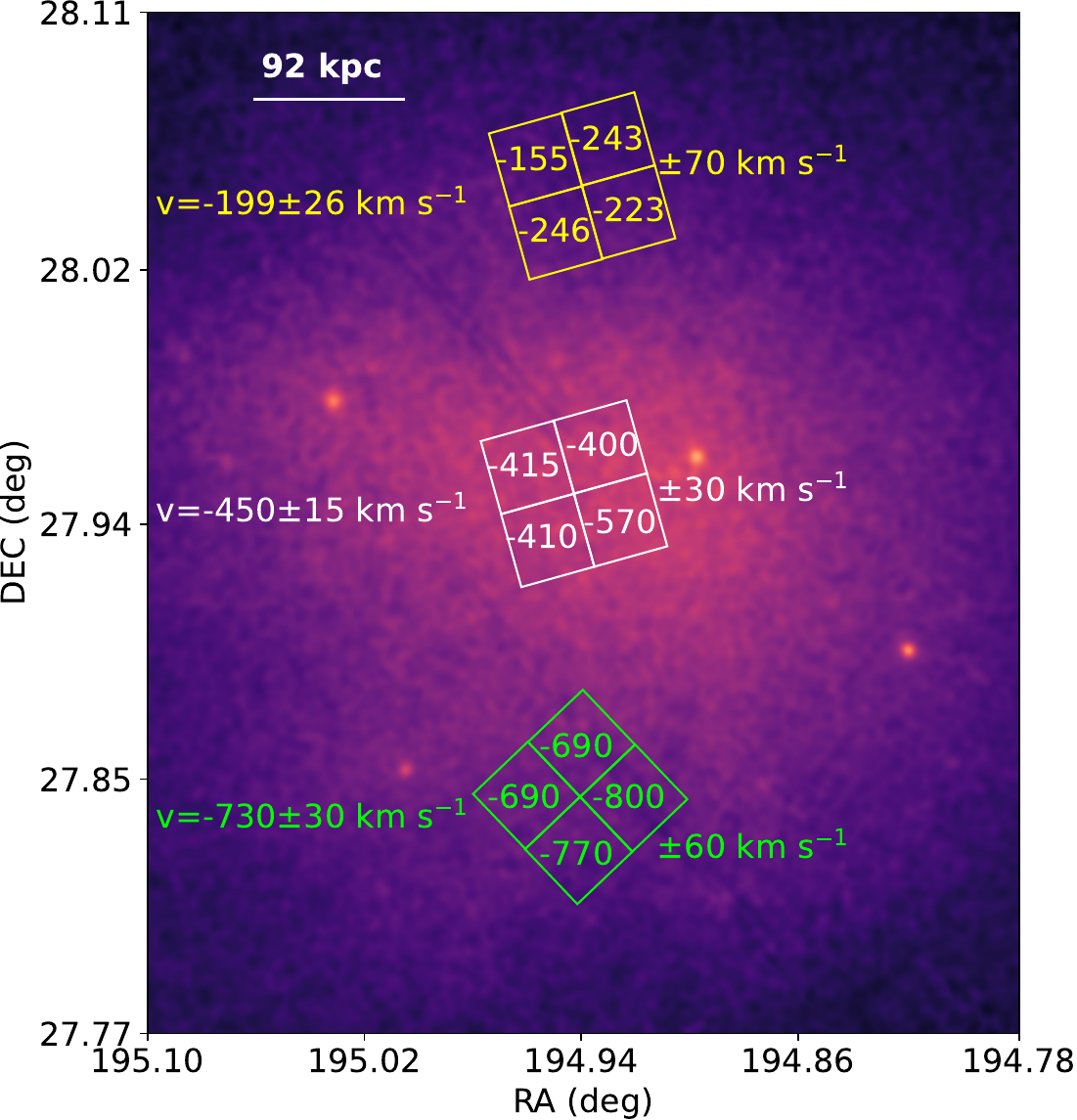}  
\caption{
Line-of-sight bulk velocities ($v_{\rm bulk}$) for the Coma cluster: northern region (yellow) and center/south (white/green; \citealt{xri25c}). 
Values are shown for the full field of view and quadrants, overlaid on the XMM-Newton X-ray image \citep{san20}. 
Quadrant points include mean uncertainties.
} \label{fig_coma_regions_vel} 
\vspace{-5mm}
\end{figure}

Best-fit parameters for the {\tt bapec} model obtained via MCMC are listed in Extended Data Table~\ref{tab_best_fit_models}. 
Convergence diagnostics confirmed robust mixing ($\tau \lesssim 1$), yielding $\gtrsim 3.8\times10^{7}$ effective samples.
The best-fit redshift corresponds to a bulk velocity of $v_{\rm bulk}=-199\pm26$~km/s and a velocity dispersion of $\sigma_{v}=167\pm39$~km/s.
These values agree with {\it XMM-Newton} results ($z=0.023$, $v=-57\pm150$~km/s; \citealt{san20}) and are notably lower than those observed in the Coma core ($-450\pm15$~km/s) and southern regions ($-730\pm30$~km/s; Fig.~\ref{fig_coma_regions_vel}), confirming significant large-scale gas motions across the cluster. 

\begin{figure}    
\centering
\includegraphics[width=0.42\textwidth]{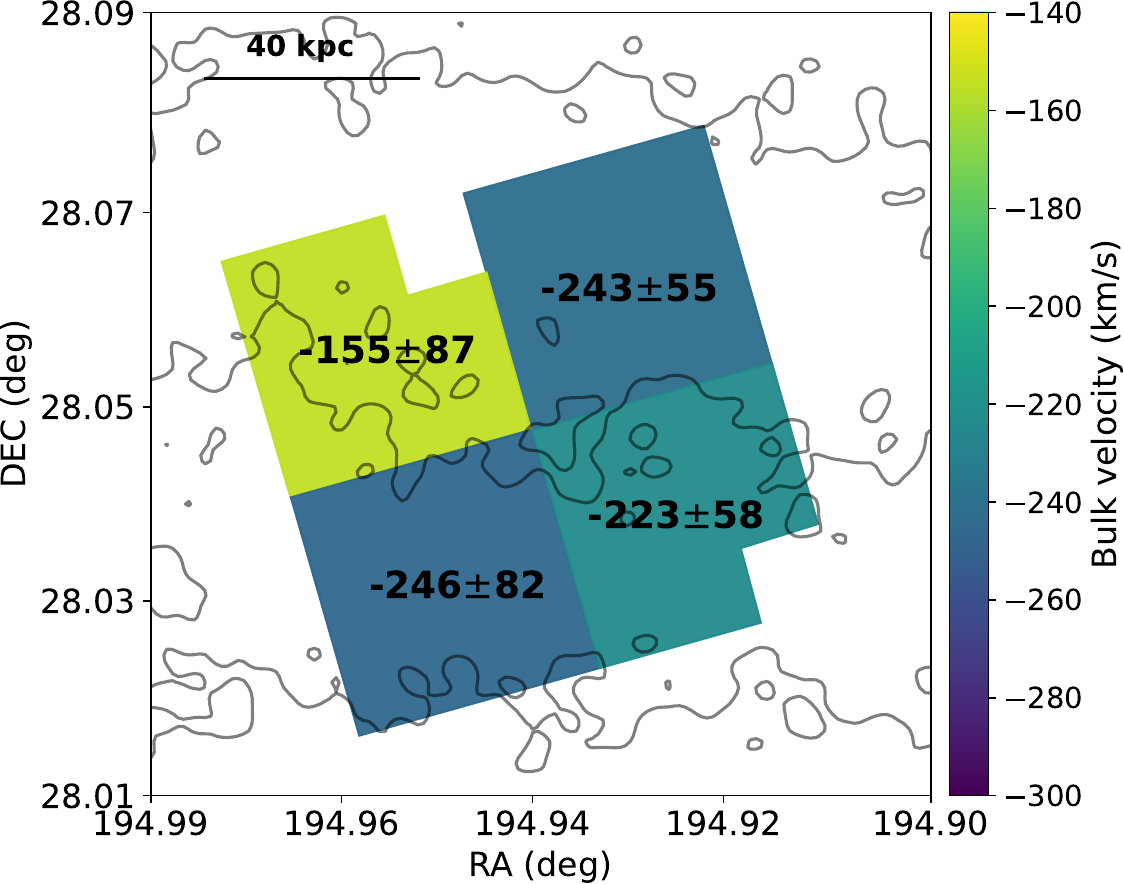} 
\includegraphics[width=0.42\textwidth]{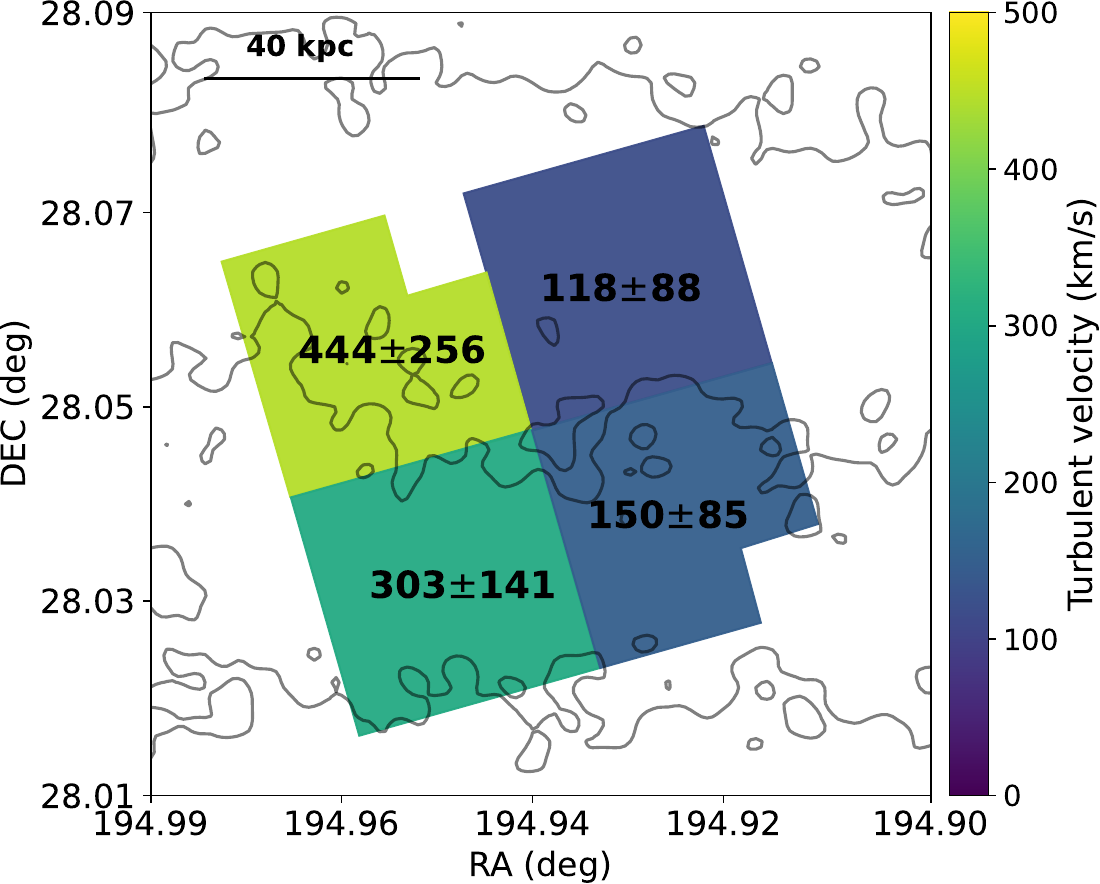}
\caption{Velocity maps obtained for the Coma cluster. 
\emph{Top panel:} Bulk velocities ($v_{\rm bulk}$) obtained from the XRISM Resolve spectra. 
\emph{Bottom panel:} Turbulent velocities ($\sigma_{v}$). 
The {\it Chandra} contours for the broadband energy range (0.5-7~keV) are included. 
}\label{fig_coma_regions_vel_contour} 
\vspace{-3mm}
\end{figure} 

 \begin{figure} 
    \centering
    \includegraphics[width=0.45\textwidth]{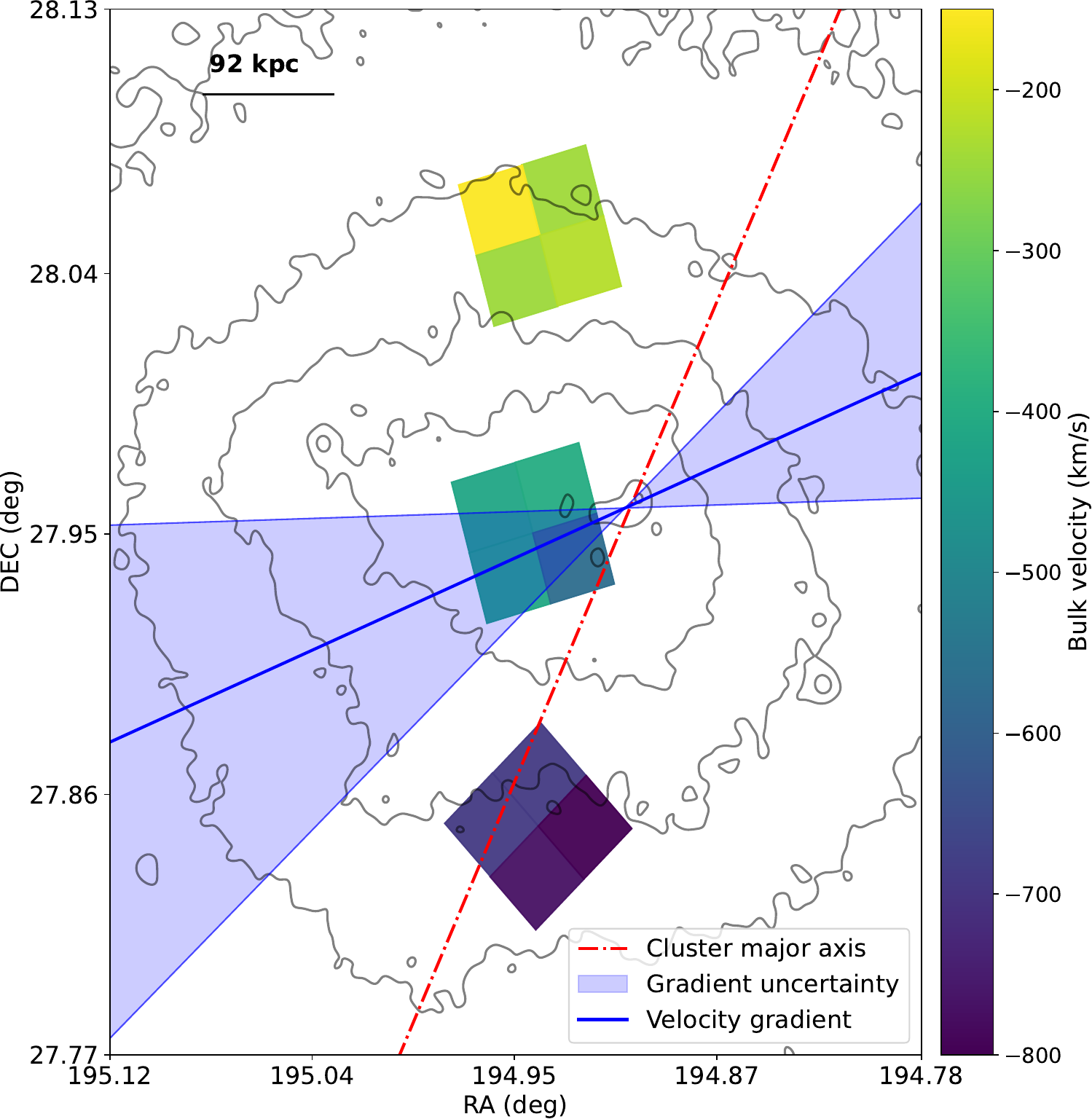}
    \caption{
Spatial distribution of Coma ICM line-of-sight velocities. 
XRISM Resolve pointings (north, center, south) are divided into quadrants (NE, NW, SE, SW) and color-coded by $v_{\rm bulk}$. 
Smoothed X-ray contours show gas density, the blue line indicates the best-fit velocity gradient (shaded region = uncertainty), and the dashed red line marks the cluster major axis (PA = 65$^\circ$). 
The plot highlights a coherent large-scale velocity shear slightly misaligned with the cluster elongation.
    }
    \label{fig:coma_velocity_gradient}
\vspace{-3mm}
\end{figure} 

\vspace{-5mm}
\subsection{Velocity structure of the hot ICM} 
To examine spatial variations in the thermodynamic and kinematic properties of the ICM, the {\it XRISM} Resolve field of view was divided into four quadrants: northeast (NE), northwest (NW), southeast (SE), and southwest (SW).
Each region contains $\sim200$ counts in the Fe~XXV–XXVI complex ($6.4$–$6.9$~keV), sufficient to constrain line centroids and measure velocity dispersions, albeit with large statistical uncertainties due to limited counts.
Spectra were fitted independently using a single-temperature {\tt bapec} model following \citet{xri25c}, with parameter uncertainties explored through MCMC sampling.
Best-fit results are listed in Table~\ref{tab_best_fit_models}, and Fig.~\ref{fig_coma_regions_vel_contour} shows the corresponding bulk and turbulent velocities.
Bulk velocities range from $v_{\rm bulk} = -246$ to $-155$~km/s, with 1$\sigma$ uncertainties of $\lesssim 100$~km/s.
Measured velocity dispersions span $\sigma_{v} = 118$-$444$~km/s, reflecting the limited photon statistics rather than intrinsic variability.
Temperatures are consistent within $kT = 7.60$-$8.55$~keV, and Fe abundances lie between $Z = 0.25$-$0.38$ (proto-solar).
Overall, no significant spatial gradients are detected, indicating that the ICM in this region is thermally and kinematically uniform within current uncertainties.

We tested the spatial uniformity of velocity, velocity dispersion, temperature, and metallicity across the four quadrants by comparing each parameter to a constant model defined by the inverse-variance weighted mean.
Reduced $\chi^2$ and $p$-value analyses, along with the largest deviations from the mean, showed no significant departures from uniformity: $0.52\sigma$ for $v_{\rm bulk}$, $0.90\sigma$ for $\sigma_v$, $0.56\sigma$ for $kT$, and $0.78\sigma$ for $Z$.
These results indicate consistency across the field, though moderate intrinsic variations may remain undetected within current uncertainties.
To further quantify possible small-scale inhomogeneities, we applied an MCMC analysis modeling the total variance as $\sigma_{\rm tot,i}^{2} = \sigma_{\rm err,i}^{2} + \sigma_{\rm intr}^{2}$.
The ICM bulk velocity is consistent across all regions, with a common mean of $v_{\rm bulk} = -226 \pm 33$~km/s and negligible intrinsic scatter ($\sigma_I < 0.01$).
The velocity dispersion, temperature, and metallicity show mild variations, with $\sigma_v = 174 \pm 55$~km/s, $kT = 8.03 \pm 0.35$~keV, and $Z = 0.28 \pm 0.03~Z_\odot$, all with $\sigma_I < 10$.
Overall, the ICM appears homogeneous on the Resolve scale, with only subtle, percent-level fluctuations consistent with the data.

\vspace{-3mm}
\subsection{Velocity structure characterization} 
To characterize the ICM velocity structure, we used the mean line-of-sight velocities from the three XRISM Resolve pointings (north, center, south). 
Projected positions relative to the cluster center (RA$_0 = 12^{\rm h}59^{\rm m}46^{\rm s}$, Dec$_0 = +27^\circ56'00''$) were converted to kpc, and the velocity field was modeled as a linear plane:
\begin{equation}
v(x,y) = v_0 + a\,x + b\,y,
\end{equation}
with $v_0$ the central velocity and $(a,b)$ the gradients along RA and Dec. 
The fit yields $a = 0.016$~km/s/kpc, $b = 4.06$~km/s/kpc, giving a total gradient $|\nabla v| = 4.06$~km/s/kpc oriented $\theta = 22.3 \pm 20.6^\circ$ east of RA, or $-42.7^\circ$ relative to the cluster major axis.
This indicates a coherent, large-scale velocity shear across the ICM, providing the first high-resolution spectroscopic measurement of bulk motions in Coma over multiple scales, extending the earlier XMM-Newton results \citep{san20}.
While the orientation is indicative given limited sampling, this direct gradient measurement offers a complementary view of ICM dynamics pending the full velocity structure function analysis in a forthcoming study. 

\vspace{-4mm} 
\subsection{Velocity of gas versus galaxies}  
To assess the local dynamical state of the ICM, we compared XRISM Resolve line-of-sight velocities with the redshifts of seven cluster member galaxies within the northern FOV \citep{mic08,den11,eis07}. 
The galaxy sample has a mean redshift $\langle z_{\rm gal}\rangle = 0.0250$ ($\langle v_{\rm gal} \rangle = 7508$~km/s) and standard deviation $\sigma_{z,\mathrm{gal}} = 0.0034$ ($\sigma_{v,\mathrm{gal}} = 1021\ $~km/s). 
A simple z-test shows the XRISM full-FOV velocity differs by only $0.68\sigma$ from the galaxy mean, indicating no strong deviation within this small sample.
To account for the large intrinsic galaxy scatter, we performed a Bayesian MCMC analysis modeling the ICM and galaxy mean velocities and the intrinsic galaxy dispersion. 
This yields a median velocity offset of $\Delta V = \mu_{\rm ICM}-\mu_{\rm gal} = -711^{+68}_{-72}$~km/s, with the $68\%$ credible interval excluding zero, indicating a statistically significant bulk motion of the ICM relative to the local galaxy population. 
These results suggest the ICM in the northern FOV is either part of a distinct kinematic substructure or exhibits coherent bulk flows, reflecting a dynamically active state.

\vspace{-4mm}
\subsection{Energy budget} 
We estimated the contribution of turbulent motions to the ICM pressure by computing the adiabatic sound speed $c_s$, the 3D Mach number $\mathcal{M}{\rm 3D} = \sqrt{3},\sigma_v/c_s$, the turbulent-to-thermal energy ratio $E{\rm turb}/E_{\rm therm} = (\gamma/2), \mathcal{M}{\rm 3D}^2$, and the kinetic pressure fraction $P{\rm turb}/P_{\rm tot} = (1+3/\gamma,\mathcal{M}_{\rm 3D}^{2})^{-1}$, where $\gamma = 5/3$ is the adiabatic index, $\mu = 0.61$ is the mean molecular weight, $m_p$ is the proton mass, and $kT$ is the best-fit ICM temperature from the spectral analysis. 
Uncertainties on all derived quantities were propagated using standard Gaussian error propagation from the measured temperature and velocity dispersion.

For the northern Coma FOV, we find $\mathcal{M}{\rm 3D} = 0.197 \pm 0.046$, $E{\rm turb}/E_{\rm therm} = 0.032 \pm 0.015$, and $P_{\rm turb}/P_{\rm tot} = 0.021 \pm 0.010$, indicating subsonic turbulence with only $\sim2\%$ non-thermal pressure support.  
These results are in excellent agreement with recent XRISM measurements: A2029 ($M_{\rm 3D} = 0.22$, $P_{\rm kin}/P_{\rm tot} = 2.6\%$), Centaurus ($M_{\rm 3D} \lesssim 0.2$, $E_{\rm turb}/E_{\rm therm} \sim 3\%$), Perseus ($M_{\rm 3D} \sim 0.1-0.3$, $P_{\rm kin}/P_{\rm tot} \sim 1-5\%$), Coma PV pointings ($M_{\rm 3D} = 0.24 \pm 0.015$, $P_{\rm kin}/P_{\rm tot} = 3.1 \pm 0.4\%$), Ophiuchus ($M_{\rm 3D} = 0.16$, $P_{\rm kin}/P_{\rm tot} = 1.4\%$), and Abell 2319 ($M_{\rm 3D} = 0.49^{+0.23}_{-0.12}$, $P_{\rm kin}/P_{\rm tot} = 11.2^{+10.6}_{-4.7}\%$). Our Mach numbers and non-thermal pressure fractions are also broadly consistent with indirect measurements based on X-ray and SZ surface-brightness fluctuations, including \cite{chu12} for Coma ($M_{\rm 3D} \sim 0.23$, $E_{\rm turb}/E_{\rm therm} \sim 3\%$), \cite{zhu18} ($E_{\rm turb}/E_{\rm therm} \sim 5\%$), \cite{dup24} ($M_{\rm 3D} \sim 0.4$, $P_{\rm turb}/P_{\rm tot} \sim 9\%$), and \cite{rom25} ($M_{\rm 3D} \sim 0.52$). 
Overall, the ICM in Coma is dynamically quiescent, with turbulence contributing modestly to the total pressure budget and subsonic motions.

\vspace{-4mm}
\section{The complex ICM dynamics of the Coma cluster}\label{sec_dyn} 
In addition to subsonic turbulence, we detect coherent bulk motions in Coma. 
Combining our northern pointing with previous XRISM measurements of the central and southern regions \citep{xri25c} reveals line-of-sight velocities of $-730$~km/s (south), $-450$~km/s (core), and $-199$~km/s (north), corresponding to a $\sim530$~km/s gradient across several hundred kiloparsecs (Figure~\ref{fig_coma_regions_vel}). 
Fitting a planar velocity model yields a gradient of $\sim4.1$~km/s/kpc oriented $22 \pm 21^\circ$ east of the RA axis, slightly misaligned with the cluster elongation, indicating a coherent velocity shear.
Subdividing each field into quadrants produces similar trends, though uncertainties limit sensitivity to small-scale variations.

Compared with other nearby clusters observed with XRISM, Coma displays distinctive kinematics. 
Its northern line-of-sight velocity dispersion ($\sigma_v = 167 \pm 39$~km/s) is comparable to relaxed systems such as the A2029 core \citep{xri25d}, yet unlike those clusters, Coma exhibits a global velocity gradient spanning hundreds of kiloparsecs. 
Despite this, the Mach number and non-thermal pressure fraction remain low, consistent with a locally relaxed ICM. 
These features are naturally explained by an off-axis merger, producing large-scale bulk flows while maintaining near-uniform thermodynamic and kinematic properties within each pointing. 
Hydrodynamical simulations and recent theoretical studies \citep{bif11,bif22,vaz25,gro25} support this scenario, showing that mergers can generate extended velocity gradients without strongly enhancing local turbulence. 
Together, these findings reinforce our observational conclusion that the Coma ICM exhibits coherent bulk motions over hundreds of kiloparsecs while maintaining relatively low turbulence and near-uniform properties within individual XRISM pointings. 

\vspace{-4mm} 
\section{Conclusions and summary}\label{sec_con}    
We have analyzed the velocity structure of the Coma galaxy cluster using new {\it XRISM}/Resolve observations of a region north of the cluster core.
The spectrum of the ICM is well described by a single-temperature model, with no significant improvement obtained from more complex descriptions.
For the entire field of view, we measure a line-of-sight bulk velocity of $v_{\rm bulk} = (-199 \pm 26)$~km/s and a velocity dispersion of $\sigma_{v} = (167 \pm 39)$~km/s.
Dividing the field into four quadrants reveals uniform thermodynamic properties and subsonic velocity dispersions, confirming the absence of strong local disturbances.
The ICM redshift is consistent with that of the cluster member galaxies detected in the same region, suggesting that the hot gas and galaxies share a common dynamical state.
From these measurements, we derive a three-dimensional Mach number of $\mathcal{M}_{\rm 3D} = 0.197 \pm 0.046$, corresponding to a turbulent-to-thermal energy ratio of $(3.2 \pm 1.5)\%$ and a kinetic pressure fraction of $(2.1 \pm 0.5)\%$ of the total ICM pressure.

When combined with previous {\it XRISM} observations of the Coma core and southern regions, our measurements reveal a coherent velocity gradient of $\sim 4.06$~km/s/kpc ($\sim 530$~km/s) across the cluster from south to north), oriented $22.3^{\circ} \pm 20.6^{\circ}$ east of the right ascension axis.
This large-scale shear, together with the uniform velocity dispersion across the pointings, provides direct evidence for an off-axis merger event that generated coherent bulk flows while leaving the local ICM relatively relaxed.
The derived subsonic motions and modest non-thermal pressure support indicate that Coma is currently in a quiescent post-merger phase.
Overall, these results establish Coma as a benchmark for studies of cluster dynamics and demonstrate the capability of high-resolution X-ray spectroscopy to map multi-scale gas motions in the ICM.
 

\vspace{-4mm}
\bibliographystyle{aa}

\begin{appendix} 
\section{Detailed Data Reduction}\label{app:data}

The {\it XRISM}/Resolve observation of the Coma northern offset region (ObsID 201114010) was performed on 2025 January 14. 
Data reduction followed standard {\it XRISM} procedures \citep{xri25a,xri25b,xri25c} using HEASoft v6.34 and XRISM {\tt CalDB} release 20241115. 
Events during South Atlantic Anomaly passages, Earth occultation, and low elevation ($<20^\circ$) from the sunlit limb were excluded, yielding a net exposure of 144.8~ks. 

The onboard $^{55}$Fe calibration source provided an energy resolution of $4.51\pm 0.02$~eV at 5.9~keV and an energy-scale offset $<0.1$~eV, corresponding to a velocity uncertainty $<10$~km/s. 
Spectra were extracted using high-resolution primary events (ITYPE=0), excluding pixel 27 due to gain instability. 
The 2–10~keV count rate was $0.0697\pm0.0007$~cts/s, totaling $\sim10100$ counts, and spectra were optimally binned to one count per channel \citep{kaa16}. 

Non-X-ray background (NXB) spectra were generated using {\tt rslnxbgen} from the Night Earth database, contributing $<8\%$ of the total counts. 
Response files were generated with {\tt rslmkrmf} and {\tt xaarfgen}, adopting the ``split-matrix'' configuration for electron-loss continuum. 
ARFs were computed from a 2--7~keV {\it Chandra} ACIS-I image (six observations, 960~ks total) with point sources removed via {\tt wavdetect}. 

To investigate spatial variations, spectra and ARFs were extracted for four quadrants (NE, NW, SE, SW), each containing sufficient counts in the Fe K complex for velocity measurements. 
PSF scattering between quadrants was modeled with {\tt xrtraytrace}, confirming negligible contamination between regions. 
This detailed data reduction ensures robust measurements of bulk velocities and turbulent line broadening across the northern Coma field.

\section{Detailed Spectral Fitting}\label{app:spectral} 
Spectral fitting was carried out in the 2--10~keV band using {\it xspec} v12.14.1 and Cash statistics \citep{cas79}. 
We modeled the ICM with three approaches:

\begin{enumerate}
    \item \textbf{Single-temperature} ({\tt bapec} v3.0.9, \citealt{fos19}), including thermal and turbulent broadening calibrated from {\it Hitomi} Perseus results \citep{hit18}. Free parameters were temperature, redshift, abundance (proto-solar, \citealt{lod09}), velocity dispersion $\sigma_v$, and normalization.
    \item \textbf{Two-temperature} ({\tt bapec+bapec}), with redshift, abundance, and $\sigma_v$ tied between components.
    \item \textbf{Continuous temperature distribution} ({\tt blognorm}, \citealt{gat22b}), assuming a log-normal temperature distribution with free median temperature, width, and $\sigma_v$.
\end{enumerate}

All models included Galactic absorption via {\tt tbabs} with fixed $N_{\rm H}=8.66\times10^{20}$~cm$^{-2}$ \citep{wil13}. 
NXB was modeled using the standard {\it XRISM} template comprising a power-law continuum and Gaussian fluorescence lines (Cr, Mn, Fe, Ni, Cu, Au), contributing $<8\%$ of counts and having negligible impact on velocity or line-width measurements \citep{xri25a,xri25b,xri25c}. 

We explored parameter uncertainties with MCMC sampling using the Goodman-Weare algorithm ({\it xspec}), with 25 walkers and $2\times10^{6}$ steps (first $5\times10^{5}$ discarded), ensuring robust estimates of bulk velocities and turbulent line widths.
Velocities were corrected to the heliocentric frame ($+24.1$~km/s). 
We adopt a flat $\Lambda$CDM cosmology with $\Omega_m=0.3$, $\Omega_\Lambda=0.7$, and $H_0=70$~km/s/Mpc. 
All reported errors correspond to 1$\sigma$ unless otherwise stated.

\begin{figure} 
\centering
\includegraphics[width=0.45\textwidth]{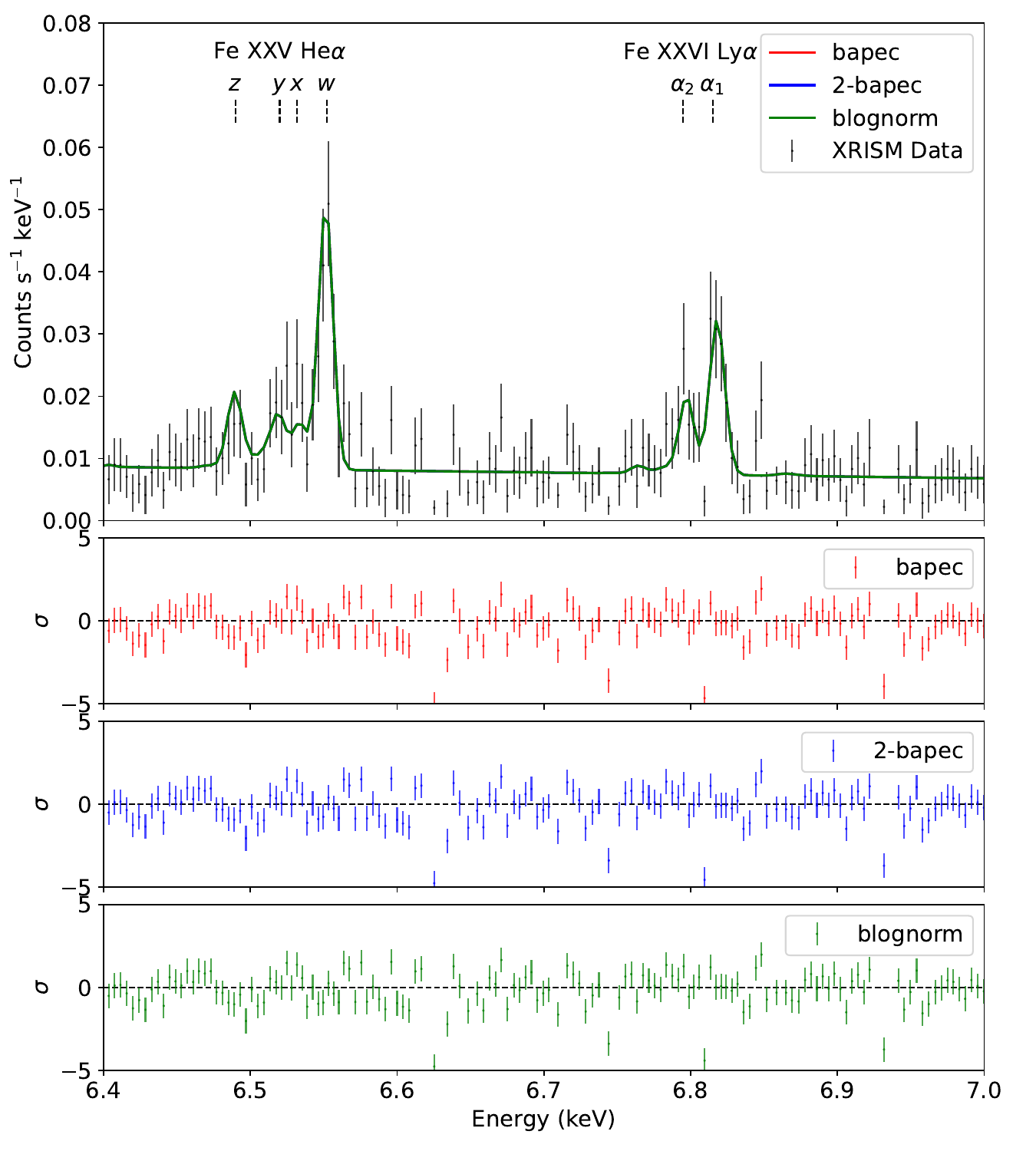} 
\caption{
XRISM Resolve spectrum zoomed on the strongest Fe K-shell lines. 
He-like triplet components are shown: resonance ($w$), intercombination ($x,y$), and forbidden ($z$); Fe~XXVI Ly$\alpha$ has $\alpha_1$ and $\alpha_2$ transitions. 
The spectrum is rebinned for clarity. 
Lower panels show fit residuals for {\tt bapec} (red), {\tt 2-bapec} (blue), and {\tt blognorm} (green). 
Minor deviations in Fe~XXVI $\alpha_2$ and Fe~XXV intercombination lines resemble those seen in A2029 and Centaurus \citep{xri25a,xri25c}.
}\label{fig_best_fit_spectra_Fe_he} 
\end{figure}

Figure~\ref{fig_best_fit_spectra_Fe_he} shows the Fe K-shell features with best-fit models and residuals.
Minor deviations in the Fe~XXVI Ly$\alpha_{2}$ and Fe~XXV intercombination lines-also seen in A2029 and Centaurus \citep{xri25a,xri25c}—likely arise from residual calibration or atomic data systematics, which are being investigated through a broader cross-comparison effort.

\end{appendix}
\end{document}